\documentstyle[12pt,epsfig]{article}
\newcommand{\lsim}{\raisebox{-0.07cm}{$\, \stackrel{<}{{\scriptstyle
\sim}}\, $}}
\newcommand{\gsim}{\raisebox{-0.07cm}{$\, \stackrel{>}{{\scriptstyle
\sim}}\, $}}
\newcommand{\beq}[1]{\begin{equation}\label{#1}}
\newcommand{\eeq}{\end{equation}}
\newcommand{\beqar}[1]{\begin{eqnarray}\label{#1}}
\newcommand{\eeqar}{\end{eqnarray}}
\newcommand{\Gl}[1]{eq.~(\ref{#1})}

\newcommand{\PRL}{Phys. Rev. Lett.\ }
\newcommand{\PL}{Phys. Lett.\ }
\newcommand{\NP}{Nucl. Phys.\ }
\newcommand{\ZP}{Zeit. f. Phys. C\ }

\begin{document}

{\flushleft  \large NTZ 26/97 }

\begin{center}
{\LARGE \bf Polarized parton distributions \\
and non-leading reggeons}

\vspace{1cm}
R.~Kirschner

\vspace*{1cm}
{\it
Naturwissenschaftlich-Theoretisches Zentrum  \\
und Institut f\"ur Theoretische Physik, Universit\"at Leipzig
\\
Augustusplatz 10, D-04109 Leipzig, Germany }

\vspace*{2cm}

\end{center}

\begin{abstract}
The small $x$ behaviour of polarized parton distributions is related to
non-leading reggeons in the framework of the high-energy effective
action. Double-logarithmic contributions to the reggeon interaction
beyond the ones included in the GLAP equation are encountered in all
polarized parton distributions. The transversity distributions of quarks
and gluons behave similar at small $x$.
\end{abstract}

\section{Parton distributions}

\vspace{1mm}
\noindent
For analyzing the small-$x$ behaviour of structure functions or parton
distributions it is appropriate to consider the $t$-channel exchange in the
corresponding forward scattering amplitutes. Owing to polarization
effects the  $P$ parity and the minimal angular momentum are the relvant
$t$-channel quantum numbers.

The unpolarized parton distributions are related to positive parity exchange
(V), the helicity asymmetry distributions to negative parity exchange (A).
 The second type of polarized parton distributions, the transverse
polarization asymmetries or transversities (T) are related to angular
momentum 1 (quark transversity) or 2 (gluon transversity) exchange.

In the deep-inelastic kinematics the exchange of two gluons at the tree
level results in a small-$x$ behaviour $x^{-a_0}$ different for the three cases:
$a_0= 1, 0, -1$ respectively for the cases V, A, T. Analogously for the
small-$x$ asymptotics of quark-antiquark exchange we have:
$a_0 = 0, 0, -1 $ for V, A, T, respectively.

In the cases V and A the quark exchange contribution can be separated
by looking into the flavour non-singlet channel. In the case T the
quark and the gluon states do not mix since they differ by angular momentum
and chirality.

\section{Non-leading reggeons}

\vspace{1mm}
\noindent
In the generic case the gluon exchange contributes to the high energy
asymtotics of the ampitude as $s^1$ and the quark exchange as $s^{1/2}$.
It is known that these leading gluon or quark exchanges can be described
in terms of perturbative reggeons. We propose to consider from this point
of view also the non-leading contributions from quark and gluon exchange,
in particular the one suppressed by one power of $s$ compared to the leading
contributions.

The leading gluon ($g_.$) or quark ($q$) exchanges can be related to
the
 transfer of spin 0 or $1/2$ through the $t$-channel. There are two
types of
leading quark exchanges distinguished by the sign of helicity or
chirality
($q_L, q_R $). The non-leading gluon ($g_{\perp }$) and quark
($q^{\prime }$)
exchanges are related to the tranfer of spin 1 and $3/2$ respectively.
 Denoting by
$\sigma_i$ the spin transferred by the reggeon $i$ the multiple
exchange of those reggeons contributes to the asymptotics $s^{a_0}$
like

\beq{1}
a_0 = 1 - \sum_i \sigma_i
\eeq

This can be considered as the extension of the known Azimov  rule
\cite{YA} to non-leading exchanges.

The contribition from gluon and quark exchanges suppressed by powers
of $s$ becomes important in particular in the small-$x$ behaviour of
polarized structure
functions. At the tree level the small-$x$ behaviour of unpolarized
structure functions (V)
is determined by two leading gluonic ($g_. g_. $) of quark
($q_L \overline q_L + q_R \overline q_R $) reggeons. The helicity
asymmetries
(A) are determined by the exchanges $ g_. g_{\perp }$ and
($q_L \overline q_L
- q_R \overline q_R $) and the transversities (T) by
($q_L \overline q_R^{\prime } + q_R \overline q_L^{\prime } + ... $)
and $g_{\perp} g_{\perp }$.

Also two or more leading reggeons in a state with corresponding
orbital angular momentum couple to the channels of the polarized parton
distributions. Such contributions enter at the higher loop level.

A systematic study of the non-leading gluonic reggeons has been performed
in the framework of the high-energy effective action \cite{KS97}. The
derivation of the
effective action from the QCD action, which we gave for the leading
reggeons, has been extended by improving the approximation keeping all
terms suppressed by one power of $s$. The reggons and their interaction
vertices with scattering quarks and gluons can be obtained by $t$-channel
factorization of corresponding $ 2 \rightarrow 2$ and $2 \rightarrow 3$
amplitudes in the multi-Regge kinematics.

The leading gluonic reggeon $g_.$ corresponds to the logitudinal
projection
of the gluon pro\-pagator in a covariant gauge or to the exchange of  the
following
combination of the transverse gauge potential $A_{\sigma },   \sigma =
1, 2 $ in the light-cone gauge:  ${\cal A}_+ = \partial_{-}^{-1}
\partial_{\sigma } A_{\sigma } $. It carries even P parity and odd  C
parity. 
Its coupling to scattering quarks and gluons conserves helicity and 
 does not
depend on helicity or transverse momenta.

One of the non-leading gluonic
reggeon carries the same quantum numbers as the leading one and its
coupling
has similar properties.

Further there are three non-leading reggeons carrying
negative P parity. They are  related to the exchange of $ {\cal
A}^{\prime } =
 i \epsilon_{ \sigma \tau} \partial_{\sigma } A_{\tau },
\sigma, \tau = 1, 2 $. These three reggeons are distinguished by their
coupling.
All three couplings conserve helicity but, unlike the leading reggeon,
depend on the sign of helicity.
One of these reggeons does not couple to quarks and another one has a
transverse momentum dependence in its coupling.

The non-leading quark reggeons have not been studied yet. However their
 main features can be obtained  by supersymmetry arguments, relying
on the similarity to the supersymmetric Yang-Mills theory,  where
gluonic and
fermionic reggeons and their couplings appear in multiplets.

\section{BFKL-type kernels and double logarithms}

The perturbative reggeons interact by emitting and absorbing gluons. The
corresponding interaction vertices differ from the original QCD vertices
and can be read off from the effective action. With the resulting
two-reggeon
interaction kernels (Fig 1a) we can write the known BFKL equation
\cite{BFKL} in the case
of two leading reggeons ($g_. g_.$) and its analogons for the other
two-reggeon channels (i,j) \cite{RK93}.

\begin{center}
\begin{figure}[thb]
\epsfig{file = 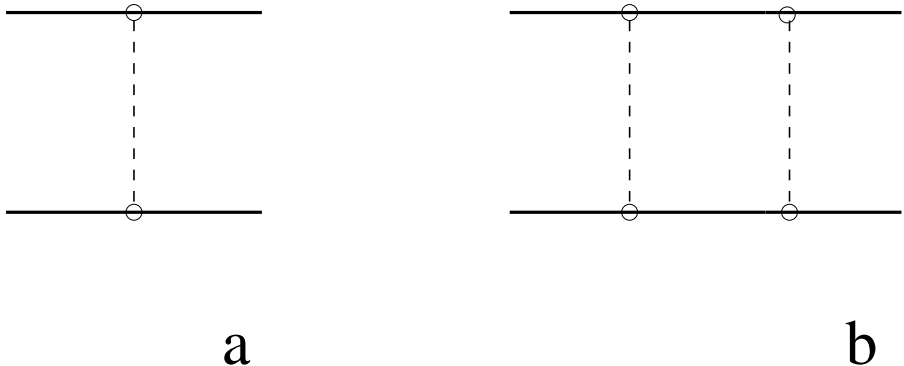, width = 120mm}
\caption{\em  a) BFKL-type kernel obtained from the effective
vertices \ \ b) Iteration of the BFKL-type equation}
\end{figure}
\end{center}

Consider a ladder graph (Fig. 1b) corresponding to the iteration of
a BFKL-type equation. We see that in all channels (i,j) corresponding
to
parton distributions the transverse momentum integrals are approximately
logarithmic in the region of strongly ordered transverse momenta,
$\vert \kappa_1 \vert \ll \vert \kappa^{\prime } \vert \ll
\vert \kappa_2 \vert $.
 This is just the contribution corresponding to the small-$x$
asymptotics of the GLAP evolution equation \cite{GLAP}.
As examples we quote the kernels for the channel $g_. g_.$ and
$q_L \overline q_L $.
\beqar{2}
K_{g_. g_. } (\kappa, \kappa^{\prime } ) =
{  2 \vert \kappa \vert^2 \vert \kappa^{\prime } \vert^2 \over
\vert \kappa - \kappa^{\prime } \vert^2 }, \cr
K_{q_L \overline q_L } (\kappa, \kappa^{\prime } ) =
{\vert \kappa \vert^2  +  \vert \kappa^{\prime } \vert^2 \over
\vert \kappa - \kappa^{\prime } \vert^2 }.
\eeqar
Here the transverse momenta are represented by complex numbers.

In the first case the only logarithmic contribution arises from the
strongly ordered region, whereas in the second case there is an additional
double-logarithmic contribution from the soft-particle intermediate state, where
$\vert \kappa_1 \vert \gg \vert \kappa^{\prime } \vert \ll
\vert \kappa_2 \vert $.

It turns out that there are double log contributions of this type in
all  channels
 corresponding to parton distributions besides of the leading channel $g_. g_.$.

In those  channels (i,j) which do not
contribute at the tree level to the parton distributions the transverse
integrals are nowhere logarithmic.
This can be seen for example in the interaction kernel of the leading
quark and anti-quark of opposite chirality ($ q_L \overline q_R $).
\beq{3}
K_{q_L \overline q_R } (\kappa, \kappa^{\prime } ) =
{2 \kappa     \kappa^{\prime }  \over
\vert \kappa - \kappa^{\prime } \vert^2 }
\eeq

In the double-logarithmic approximation the contributions from the
kernels non-logarithmic in the transverse momenta are neglected. Then
the
BFKL-type equations reduce to simpler equations. In the leading gluon
channel ($g_. g_.$) dominating the unpolarized flavour singlet
parton distribution
we arrive just at the GLAP equation with the kernels
approximated by its leading terms at small $x$. However in
the other cases of parton distributions, due to the soft-particle
intermediate states, we obtain a non-linear equation  \cite{KL82}
 given schematically in Fig. 2. Here the amplitude enters with a lower
cut-off $\kappa$ in the transverse momenta. The Born term of this
equation is directly obtained from the BFKL-type kernel $K(\kappa,
\kappa^{\prime } )$ at $\vert\kappa \vert \ll \vert \kappa^{\prime }
\vert $.

\begin{center}
\begin{figure}[thb]
\epsfig{file = 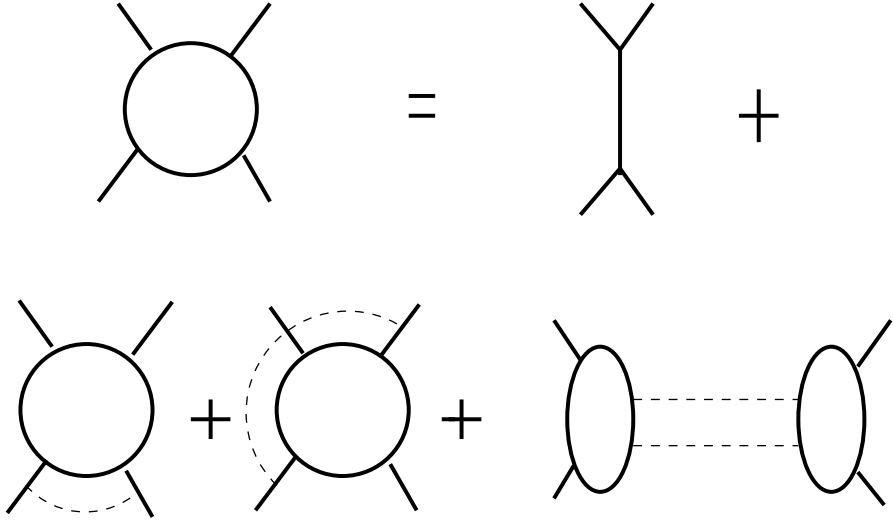, width = 120mm}
\caption{\em  The equation for the double log amplitudes}
\end{figure}
\end{center}

We emphasize once more that in these cases the small-$x$ asymptotics
of the
GLAP equation does not account for all perturbative contributions even
in the double logarithmic approximation. The naive extrapolation of the
GLAP equation to the small-$x$ region
does not represent the predictions of perturbative QCD even at the
one-loop level as far as the
dependence on $x$ is concerned. On the other hand the GLAP equation
describes correctly the $Q^2$-dependence in the small-$x$ region if the
anomalous dimensions are improved by the resummation which can be
extracted from the result of the non-linear double logarithmic eqation.
The effect of including the resummation depends of whether the input
dependence on $x$ is compatible with the non-linear equation or not
\cite{BVJap}.

\section{The transversity distributions at small $x$}

\vspace{1mm}
\noindent
The transversity distribution of quarks contributes  to the structure
function $ h_1(x, Q^2) $, which appears e.g. in the Drell-Yan pair
production with both incident particles  polarized. It is related to the
quark scattering amplitude with helicity flip, where quark and anti-quark
of opposite chirality or parallel helicity are exchanged in the
$t$-channel \cite{RaSo,JaJi,WS}.

The transversity distribution of gluons contributes to the structure
function $F_3^{\gamma } (x, Q^2 ) $, which enters the deep inelastic
scattering cross section on a polarised photon. It is related to the
amplitude of gluon scattering with helicity flip \cite{ChS,AhR,JaMa}.

The result of our analysis \cite{KMSS,EKS} is that both parton
distributions
of type $T$ behave quite similar at small $x$. In the above discussion
we have provided some  arguments why this should be expected.

The GLAP evolution kernels or anomalous dimensions \cite{BFKL2,AM}
coincide at one loop up to colour factors, which can be understood
from supersymmetry arguments \cite{BFKL2}.  Summing the double
logarithmic
contributions results essentially in the same answer up to some
colour factors. The small-$x$ asymptotics of the gluonic or quark
transversity is determined by the solution of the non-linear equation
Fig.
2 for the colour singlet positive signature channel in terms of partial
waves ($\omega = 1+ j $ ) ,
\beq{4}
 f_0^{+} (\omega ) =  4 \pi^2 \omega
\left( 1 - \sqrt { 1 - {4 \alpha_S C \over \pi \omega^2 }
 ( 1 - {1 \over 2 \pi \omega } f_8^{-} (\omega ) ) } \right)
\eeq
This solution is give in terms of the solution for the colour
octet channel of the negative signature.
\beqar{5}
f_8^{-} (\omega ) = 4 \pi \alpha_S N { d \over d \omega } \ln
 \left ( \exp ({\omega^2 \over 4 \overline \omega^2 } ) \ \ \ 
{\cal D}_{p} ({\omega \over \overline \omega }) \right ), \ \ \ \ \ 
\overline \omega^2 = {\alpha_S N \over 2 \pi }.
\eeqar
$N$ is the number of colours. In the gluonic case we have $C = N$
and in the quark case $C = {N^2 -1 \over 2 N}$.
${\cal D}_{p} (z) $ is the parabolic cylinder function. Its index $p$
 is a ratio of colour factors which is $p = 2$ in the gluonic
case and $p= - {1 \over 2 N^2} $ in the case of quark transversity.

The $x$ dependence of the parton transversity distributions is obtained
by inverse Mellin transformation from the partial waves \Gl4 and the
leading
small $x$ behaviour is $x^{1 - \omega_0^{+} }$ where $\omega_0^{+}$ is
 the rightmost singularity of the partial wave \Gl4 in $\omega $.

In this approximation also the resummed anomalous dimension $\nu (j)$
near $j = -1$ is obtained in terms the solution of the non-linear
equation $f_0^{+} (\omega ), \omega = j +1 $.

\beq{6}
\nu (-1 + \omega ) = {1 \over 8 \pi^2 }
f_0^{+} (\omega )
\eeq

The recently calculated two-loop contributions to the anomalous
dimension \cite{HKK,WV,KM} of the quark transversity are compatible with this
result\footnote{ In \cite{KMSS} there are misplaced factors of 2
in the expression for $f_8^+ (\omega )$, compare \Gl5.} near $j = -1$.

The transversities provide two examples for the situation that the
amplitudes
of deep-inelastic scattering, dominated in the leading $\log Q^2 $
approximation at small $x$ by non-leading reggeons   ($g_{\perp}
g_{\perp } $ or
$ q_L \overline q_R^{\prime }+ ... $), receive also contributions  from
the
leading reggeons ($ g_. g_. $ or $q_L \overline q_R + q_R \overline q_L $).
These exchanges couple weakly via higher powers of $\alpha_S$ not
accompanied by
large logarithms. However, they have a stronger small-$x$ asymptotics.

From the solutions of the BFKL-type equations we obtain for the
small-$x$ behaviour  of these contributions  $x^{-a }$, where
in the case of the quark transversity \cite{KMSS}
\beq{7}
a_q = {\alpha_S (N^2-1) \over 8 N \pi } \Omega (1,0) = 0
\eeq
and in the case of the gluon transversity \cite{EKS}
\beq{8}
a_g = 1 + {\alpha_S N \over 4 \pi } \Omega (2,0) =
1 + { 4 \alpha_S N \over \pi} (\ln 2 - 1).
\eeq

$\Omega (n, \nu) $ denotes the eigenvalue function of the BFKL
equation \cite{L86}
depending on the conformal spin $n$ and the anomalous dimension $\nu $.
In amplitudes
related to unpolarized scattering the case $n = 0 $ is relevant
and  $\Omega (0,0) $ leads to the known expression for the BFKL pomeron
intercept.
$n$ plays the role of the transferred orbital angular momentum.



\vspace{1cm}
\noindent
{\Large \bf  Acknowledgements}

\vspace{1mm}
\noindent
I thank the organizers of the workshop for invitation. This contribution
is based in part on the joint work with B. Ermolaev and L. Szymanowski,
which is supported by Volkswagen-Stiftung.


\end{document}